\documentclass[referee]{aa} 
\usepackage{graphicx}
\usepackage{txfonts}
\begin{document}
   \title{Photometric observations of Rosetta target asteroid 2867 Steins}

   \subtitle{}

   \author{Paul R. Weissman
          \inst{1}
          \and
          Stephen C. Lowry\inst{2}
          \and
          Young-Jun Choi\inst{1}
          }

   \offprints{P. Weissman}

   \institute{Science Division, Jet Propulsion Laboratory, 4800 Oak Grove Drive, 
              MS 183-301, Pasadena, CA 91109  USA\\
              \email{paul.r.weissman@jpl.nasa.gov,  young-jun.choi@jpl.nasa.gov}
         \and
             School of Mathematics and Physics, Queen's University Belfast, Belfast, 
		BT7 1NN, UK  \\
             \email{s.c.lowry@qub.ac.uk}
             }

   \date{Received date; accepted date}

   \abstract{Asteroid 2867 Steins is one of two flyby targets of ESA's 
	International Rosetta Mission, launched in March, 2004. We obtained
	CCD observations of Steins on April 14-16, 2004
	at Table Mountain Observatory, California, in order to characterize the 
	asteroid physically, information that is crucial for planning the Steins flyby. 
	This study includes the first detailed analysis of the physical properties of Steins from time-series 
	$R$-filter data along with $V$- and $I$-filter photometric measurements.
	We found a mean $R$-filter absolute magnitude of $12.60$~$\pm$~$0.02$ (for G=0.15), 	
	corresponding to a mean radius of $3.57$~$\pm$~$0.03$~km assuming an S-type reflectance of 	
	0.20, or $2.24$~$\pm$~$0.02$~km assuming an E-type reflectance of 	
	0.40 (and G=0.40).  The observed brightness range of $0.29$~$\pm$~$0.04$ magnitudes suggests a lower limit 
	on the axial ratio, $a/b$, of 1.30.
	We determined a synodic rotation period of $6.048$~$\pm$~$0.007$ hours, assuming a 
	double-peaked lightcurve. 
	We fitted the available $R$-filter photometry over the phase angle range
	of 11.08--17.07 degrees and found best-fit phase function parameters of 
	G=$0.46^{+0.32}_{-0.20}$, and H=$12.92^{+0.22}_{-0.17}$.
	Derived colour indices for the asteroid are ($V-R$) = $0.58$~$\pm$~$0.03$, and  
	($R-I$) = $0.44$~$\pm$~$0.03$.
	These values are consistent with, though slightly redder than Hicks et al. (IAUC 8315). 
	Barucci et al. (\cite{barucci}) identified Steins as an E-type
	based on visual and near-infrared spectra, but if that is correct,
	then it is an unusually red E-type asteroid.
 	}

   \titlerunning{Physical properties of asteroid 2867 Steins}
   \maketitle
%

\section{Introduction}

Asteroid 2867 Steins is one of two flyby targets of the European Space Agency's 
International Rosetta Mission, launched on March 2, 2004.  Rosetta will rendezvous 
with periodic comet 67P/Churyumov-Gerasimenko in 2014.  En route to 
the comet, Rosetta will fly by two main belt asteroids, 2867 Steins 
on September 5, 2008 and 21 Lutetia on July 10, 2010.  
As the targets were changed
due to the Rosetta launch delay from 2003 to 2004, the new target comet and flyby asteroids
have little observational data available (with the exception of Lutetia). 
Thus, new studies like this one are needed to 
build up a detailed picture of the physical and surface characteristics of 
asteroid Steins. 

Asteroid 2867 Steins is located in the inner main belt at a semi-major 
axis of 2.363 AU.  Orbital elements for the asteroid are given 
in Table \ref{orbital_elements}.  Its absolute $R$-band magnitude of 12.67 from 
unpublished measurements (Hicks et al. IAUC 8315) corresponds to a mean 
diameter of 6.9 km if it is a typical S-type with albedo = 0.20 or 4.9 km 
for a E-type object with albedo = 0.40.  
Hicks et al. found $BVRI$ colours consistent with an S-type taxonomic 
classification, though they could not rule out a D-type object.  However, 
Barucci et al. (\cite{barucci}) 
classified 2867 Steins as an E-type asteroid based on visual and 
near-infrared spectra taken in January and May, 2004.  
E-type asteroids are similar to enstatite achondrite meteorites, one of 
the more thermally processed classes. 

Polarization curves at wavelengths of $R$ and $V$ passbands were obtained
by Fornasier et al. (\cite{fornasier}) from observations with ESO's VLT 
telescope, and suggest an albedo of $0.45$~$\pm$~$0.10$. 
Unfortunately, with this method
there is a large scatter in the albedo/polarimetric-slope relation that 
will inevitably lead to a large uncertainty in the implied albedo.
Preliminary analysis of observations of Steins with the Spitzer Space Telescope, 
combined with visual data,
suggest a somewhat lower albedo of $0.30-0.40$ (Lamy~et~al. \cite{lamy}).

We observed 2867 Steins in April, 2004 to characterize the 
asteroid physically.
Knowledge of the asteroid's size, shape, 
rotation period, and spectral behaviour are crucial in planning the 
science observations of Rosetta.  In addition, lightcurves obtained 
at multiple apparitions can be used to derive a full 
three-dimensional shape model and rotation pole orientation, again
very important for planning of the Steins flyby.  
This was done with great success by Kaasalainen et al. (\cite{kaasalainen}) for asteroid 25143 Itokawa, 
target of the Hayabusa sample return mission. Comparison of spacecraft observations with 
results derived from telescopic observations also provides critical 
ground truth for remote observing and data analysis techniques.

This paper is organized as follows. Section \ref{observations} describes the 
observational configuration and strategy employed for this programme.
Section \ref{analysis} describes the analysis techniques that were
applied to the data, and the results on the bulk physical and surface
properties are also presented. A summary of the results and main conclusions
is given in the final section.\\


\section{Observations of 2867 Steins}
\label{observations}

Asteroid 2867 Steins was observed with the 0.6 meter, 
Ritchey-Chretien telescope at the Table Mountain Observatory (TMO), 
near Wrightwood, California on April 14-16, 2004 (UT).  The observing 
geometry of the asteroid on each date is listed in Table \ref{observing_geometry}.  
Images were obtained with the facility 
CCD camera, using a Photometrics $1024 \times 1024$ pixel thinned and back-illuminated 
CCD, mounted at the Cassegrain focus of the f/16 telescope.  The pixel 
scale was 0.52$^{\prime\prime}$/pixel and the total field-of-view was 
$8.9^{\prime} \times 8.9^{\prime}$.  
For all images the telescope was tracked at the asteroid's predicted 
rate of motion based on an ephemeris from JPL's Horizons 
system (Giorgini and Yeomans \cite{giorgini}).

The asteroid was located slightly past opposition and at a northern 
declination of +20$^{\circ}$, allowing observations below two air masses 
for about six hours per night.  Observations were conducted using the 
Johnson/Kron-Cousins $VRI$-filter set.  A total of 38 $R$-filter images of 
the asteroid field were obtained on night 1, 36 on night 2, and 39 on 
night 3.  Typical exposure times were 360-600 seconds.  Additionally, on 
night 2, one $V$-filter and one $I$-filter image each were obtained as 
part of an {\it R-V-R-I-R} sequence.  Each night's observations included 
twilight sky frames in the $R$-filter (plus the $V$ and $I$ filters on night 2), 
bias (zero exposure) frames, 
and Landolt fields (Landolt \cite{landolt}) imaged over a range of 
air masses.  Observing conditions on nights 1 and 3 were 
photometric while night 2 suffered from some high cirrus.  Because 
of this, calibration frames of the Steins star fields for night 2 
were obtained on a later run at TMO under photometric conditions.  
The Landolt fields used were PG1047+003, PG1323-086, and PG1528+062 
(Landolt \cite{landolt}).

\section{Data reduction and analysis}
\label{analysis}

All images were bias-subtracted and flat-fielded, and other instrumental 
artifacts, such as cosmic rays and bad rows/columns were removed in the 
standard manner.  Standard aperture photometry was applied to the images 
using an aperture $\sim$3 times the maximum measured seeing each night.  
Image processing was performed using the Image Reduction and Analysis 
Facility (IRAF) (Tody \cite{tody86}; \cite{tody93}).  

  \subsection{Shape, size, and surface colour}
  \label{Size_and_shape}


A relative asteroid lightcurve was extracted from the imaging for each night by comparing 
the asteroid's $R$-filter brightness with that of non-varying field stars in the 
same images.  
The observed rotational lightcurve was asymmetric, and the observed full brightness range was
$0.29$~$\pm$~$0.04$ magnitudes. This implies an axial ratio, $a/b$, of 
$\geq$~$1.30$~$\pm$~$0.04$, using $a/b \geq 10^{0.4 \Delta m}$,
where $a$ and $b$ are the semi-axes of the asteroid and $\Delta m$ is the range 
of observed magnitudes. This simple model assumes a bi-axial ellipsoid shape 
and uniform surface albedo, and ignores phase effects.
Zappal\`{a} et al. 
(\cite{zappala}) analysed the relationship between lightcurve amplitudes
and phase angles for a subset of known asteroid classes.  They showed that the
lightcurve amplitude increases with phase angle, and that the degree of change
is dependent on the taxonomic class of the asteriod. More specifically,
\begin{equation}
\label{Zappala_eq}
\Delta m(0^{\circ}) = \frac{\Delta m(\alpha)}{1+ \kappa \alpha}
\end{equation}
where $\Delta m(0^{\circ})$ and $\Delta m(\alpha)$ are the lightcurve amplitudes
at $0^{\circ}$ and $\alpha^{\circ}$. The constant $\kappa$ [deg$^{-1}$] has 
values of 0.030, 0.015, 0.014, and 0.013 for S, M, R, and C asteroid taxonomic types, 
respectively. Taking the $\kappa$ value for S-types and applying 
equation \ref{Zappala_eq} gives $\Delta m(0^{\circ})$ = 0.188 and a corrected axial ratio
of 1.19. Unfortunately, no $\kappa$ value is available for E-type asteroids
therefore we use the S-type value based on the very red colour indices of 2867 Steins (see below).

The calibrated apparent magnitudes from our Table Mountain run on April 
14-16, 2005 are listed in Table \ref{table_phot} (see the online version of this
paper). 
The mean calibrated $R$-filter 
apparent magnitude, or the apparent magnitude at lightcurve mid-point is $16.80$~$\pm$~$0.02$. 
If we assume an S-type phase-slope parameter G of 0.15 in the standard HG system of 
Bowell et al (\cite{bowell}), 
as used by Hicks et al. (IAUC 8315),
we obtain a mean absolute $R$-filter magnitude of $12.60$~$\pm$~$0.02$.  This is
brighter by 0.07 magnitudes than the Hicks et al. measurement, but consistent 
at the 2$\sigma$ level. Alternatively, if we assume 
a typical G value of 0.40 for an E-type asteroid, we obtain a mean absolute 
$R$-filter magnitude of $12.86$~$\pm$~$0.02$.

Assuming a typical S-type albedo of 0.20 (and S-type G value), our mean apparent $R$-filter
magnitude implies a mean effective radius of 
$3.57$~$\pm$~$0.03$~km or semi-axes of $3.42$~$\pm$~$0.03$ and 
$4.06$~$\pm$~$0.03$~km if the phase-angle-corrected
lightcurve magnitude range is considered. 
Assuming a typical E-type albedo 
of 0.40 (and E-type G value of 0.40), our mean apparent
magnitude implies a smaller mean effective radius of 
$2.24$~$\pm$~$0.02$~km with semi-axes of $2.14$~$\pm$~$0.02$ and 
$2.54$~$\pm$~$0.02$~km, again using the corrected axial ratio of 1.19.
We use the size-magnitude relationship of Russell (\cite{russell}).


To remove the confounding effects on the derived colour indices due to possible projected 
area variation with rotation, we compared the $V$ and $I$ apparent magnitudes
with interpolated $R$-filter measurements from Table \ref{table_phot} (included in the
online version of this paper). We took the 
average of the two bracketing $R$-filter data points for each of the $V$ and $I$ 
measurements, and measured the difference to get the ($V-R$) and ($R-I$) colour indices. 
Our quoted ($V-I$) is the summation of these two colour indices and not the 
difference between the actual $V$ and $I$ apparent magnitudes. 

\begin{figure}[t]
\begin{center}
\resizebox{\hsize}{!}{\includegraphics{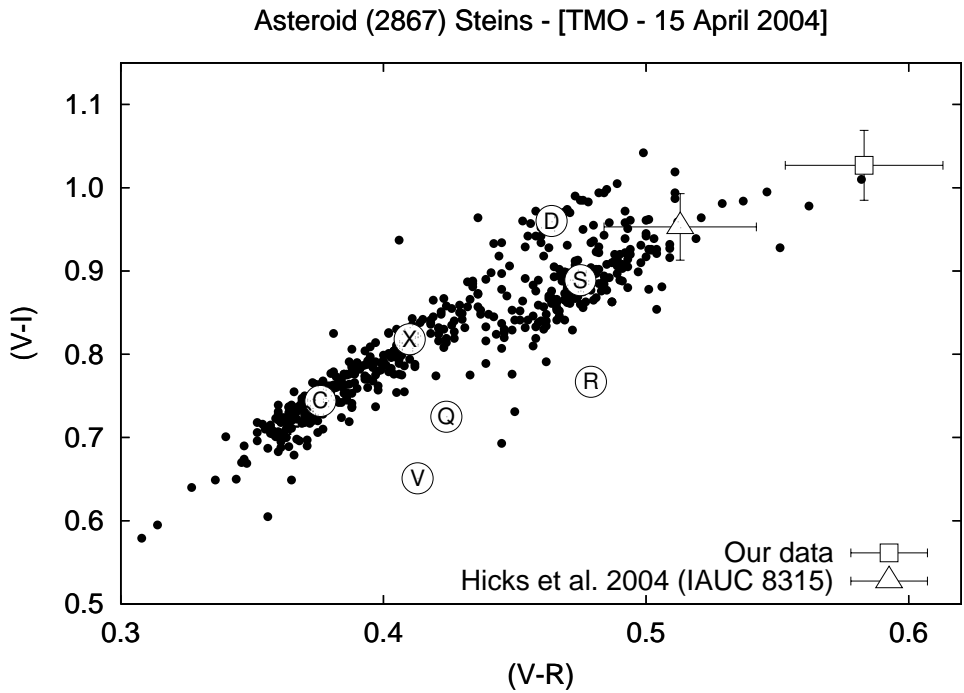}}
\caption[Caption]
{Comparison of our colours with Hicks et al. (IAUC 8315), and ECAS data from Zellner et al. (\cite{zellner}).
The positions of the main Tholen classes are marked (Tholen \cite{tholen}).
Our colour data is consistent with Hicks et al. at the 1.2$\sigma$ level, but are much redder
than typical E-types (i.e. X-type) bodies. Barucci et al. (\cite{barucci}) assign an E-type
classification to Steins due to the presence of a 0.49~$\mu$m absorption feature in their 
spectrum.
}
\label{color}
\end{center}
\end{figure}

\begin{figure}[t]
\begin{center}
\resizebox{\hsize}{!}{\includegraphics{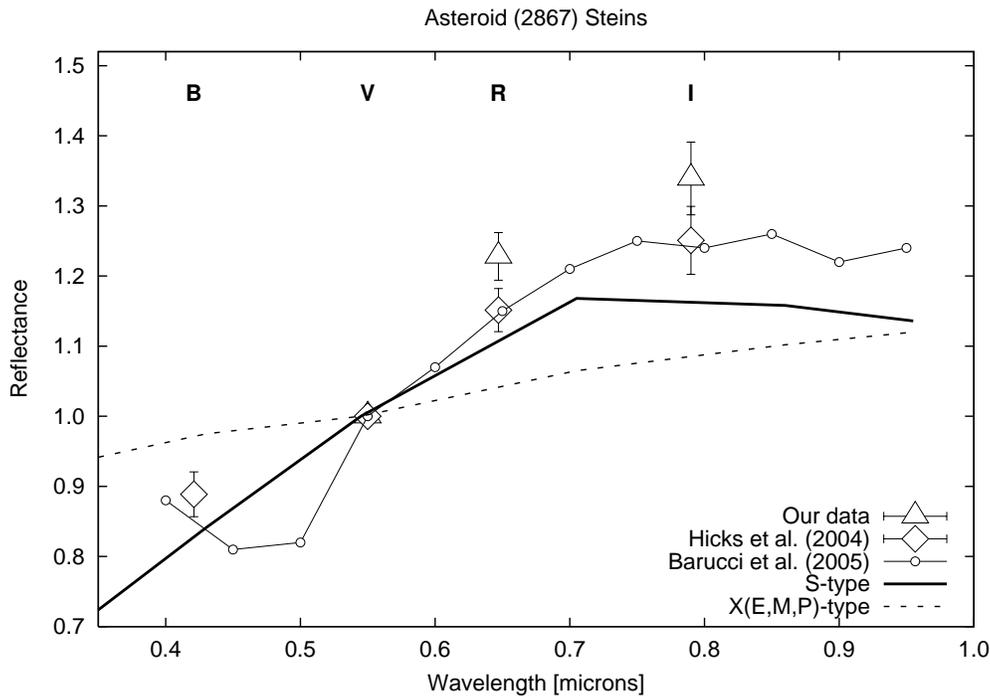}}
\caption[Caption]
{Here we compare the available visual broad-band photometry and spectroscopy for Steins. 
We include the reflectance spectrum from Barucci et al. (\cite{barucci}), which has been sampled at 0.05 $\mu$m intervals, 
and normalized at 0.55 $\mu$m. The broadband colours have been converted to reflectances and normalized
at 0.55 $\mu$m (i.e. at $V$-filter wavelengths). Our $VRI$ data is included, along with $BVRI$ photometry from Hicks
et al. (IAUC 8315).
One can see that the Barucci et al. spectrum agrees
very well with the Hicks et al. data set. Our data is slightly redder than the rest, but still in 
agreement within the photometry uncertainties and the noise level of the spectrum. We have also included
mean reflectance spectra for the Tholen X- and S-types. 
The photometry and spectroscopy all show that the spectrum is much redder than a typical E-type, 
though there are a few E-type outliers that approach the redness of the Steins spectrum.
}
\label{Barucci_comp}
\end{center}
\end{figure}

The interpolated colour indices are given in Table \ref{table:4}, along with the 
corresponding $R$ magnitude, and the values found by Hicks et al.~(IAUC 8315).  
Our values are somewhat redder than Hicks et al.~but consistent at the $1.2\sigma$ level. 
In Figure \ref{color} we compare these colours with similar colour indices for 
other asteroids, derived from ECAS data (\cite{zellner}) using the transformation method
discussed in Dandy et al. (\cite{dandy}).
Based on our colours we find that the visible colours for this asteroid are much redder
than typical E-type asteroids. As noted above, Barucci et al.~(\cite{barucci})
report an E-type classification for Steins based on visual and near-infrared spectra,
in particular the presence of a 0.49~$\mu$m absorption band.  
New preliminary colour data from Weissman et al. (personal communication), based on 
observations from Cerro Tololo in August 2005, give results similar to both Hicks et al. and 
our results presented here.  Thus, all available visible colours of 2867 Steins agree. 
We compare our results with those of Hicks et~al. 
(\cite{hicks}) and the Barucci et al. (\cite{barucci}) spectrum in Figure \ref{Barucci_comp}.  
All three data sets of Steins agree on its strong red colour.
If Steins is an E-type asteroid, then it is unusually red.

The differences in our colours and those by Hicks et al. (\cite{hicks}) may be indicative of surface 
inhomogeneities, resulting in areas of differing spectral slopes, perhaps from 
shifting of surface regolith. Such a mechanism could explain
the fascinating surface characteristics of asteroid Itokawa as observed by the Hayabusa 
asteroid-rendezvous spacecraft (Fujiwara et al.~\cite{fujiwara}; Abe et al.~\cite{abe}).  
This issue could be addressed by obtaining rotationally resolved spectra of 2867 Steins.  
Otherwise, it must await the examination of Steins by the Rosetta spacecraft in 2008.

  \subsection{Photometric phase function}
  \label{photometric_phase_function}

We used the available $R$-filter photometry to assess the phase-angle
variation of the asteroid's brightness. Our data are combined with the Hicks et al. data 
(IAUC 8315) to perform a fit of the photometric phase function in terms of the HG formalism.
The Hicks et al. data and our data were taken when Steins was at average phase angles of 
11.07 and 17.07 degrees, respectively. For the fit we use Hicks et al's. mean apparent 
magnitude of $16.53$~$\pm$~$0.02$ and convert to `reduced' magnitude R(1,1,$\alpha$), i.e. the apparent
magnitude has been scaled to unit heliocentric and geocentric distances. 
The reduced magnitude is $13.36$~$\pm$~$0.02$. We do the same for our mean $R$-filter 
apparent magnitude of $16.80$~$\pm$~$0.02$, and the resulting reduced magnitude is $13.51$~$\pm$~$0.02$. 

\begin{figure}[t]
\begin{center}
\resizebox{\hsize}{!}{\includegraphics{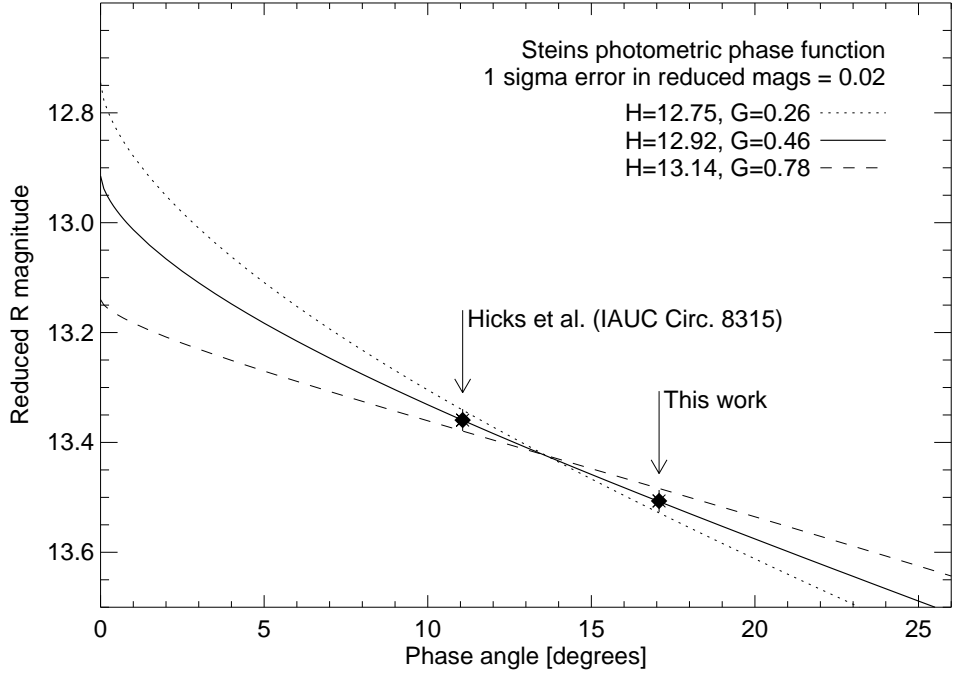}}
\caption[Caption]
{
We use the available $R$-filter photometry for a preliminary assessment of the phase-angle
variation of the asteroid's brightness in terms of the HG formalism. 
The best-fit phase function parameters and associated 1$\sigma$ uncertainties are 
G=$0.46^{+0.32}_{-0.20}$, and H=$12.92^{+0.22}_{-0.17}$. Opposition-surge coverage would 
be useful along with observations at large phase angles in order to refine this measurement.
}
\label{Phase_function_fig}
\end{center}
\end{figure}

The best fit photometric values are G=0.46 and H=12.92. When the 1$\sigma$ photometric uncertainties
in the mean apparent magnitudes are considered we find that G values of 0.26 and 0.78 fit the data
just as well. The corresponding H values are 12.75 and 13.14, respectively, and the fits are shown 
graphically in Figure \ref{Phase_function_fig}. The accepted phase function parameters
and 1$\sigma$ uncertainties are therefore G=$0.46^{+0.32}_{-0.20}$, and H=$12.92^{+0.22}_{-0.17}$.
Although there is no phase coverage within the opposition-surge region at small phase angles
we still prefer to fit for G rather than simply applying a linear phase law that is 
inappropriate for asteroidal bodies. 

A shallow slope parameter of G=0.46 is more associated with E-type asteroids than S-type. 
Because of the large error bar we cannot put too much weight on our derived slope parameter as a means of 
distinguishing whether or not the asteroid is either type. At the very least, this would require 
refinement of our phase-slope measurement through opposition surge coverage along with observations at 
large phase angles. 
Nevertheless, this result will aid in better assessment of
the brightness of Steins for future observational planning, which needs to take place 
at a wide range of observational geometries, allowing for the full 3-D shape and orientational
modelling to take place. Taken together, these data will also allow a full Hapke-type analysis to be performed 
(e.g. Lederer et al. \cite{lederer}), providing detailed information on Steins' surface regolith.

For completeness, we compute the corresponding effective radius using this fitted H value and 
for assumed typical geometric albedos of both S-type and E-type asteroids.
For a typical S-type albedo of 0.20, 
the absolute magnitude corresponds to an effective radius of $3.08^{+0.25}_{-0.30}$ km, and for 
an E-type albedo of 0.40, 
the absolute magnitude corresponds to an effective radius of $2.18^{+0.18}_{-0.21}$ km.

   \subsection{Rotational properties}
   \label{rotation_properties}
   
\begin{figure}[t]
\begin{center}
\resizebox{\hsize}{!}{\includegraphics{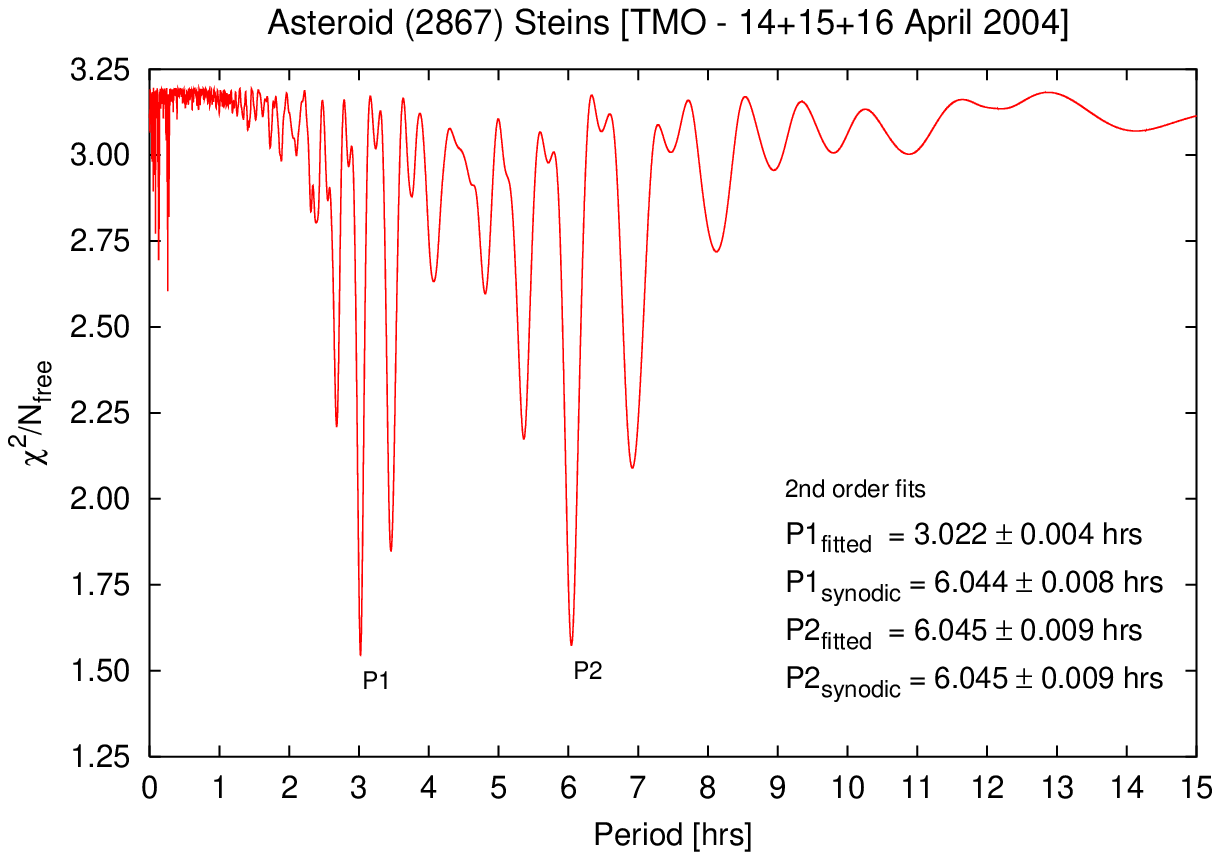}}
\resizebox{\hsize}{!}{\includegraphics{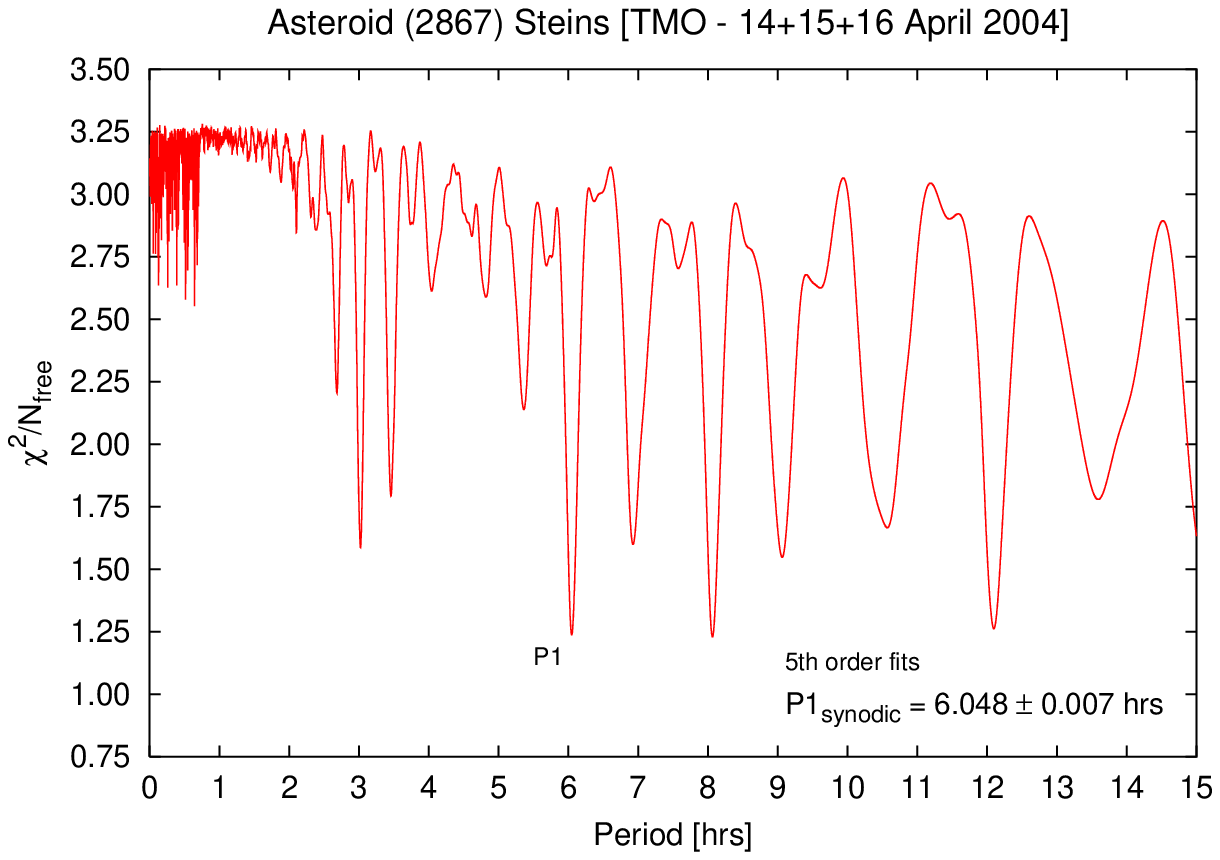}}
\caption[Caption]
{\textit{Upper panel:} Resulting periodogram from 2nd order Fourier fits to the relative magnitudes.
We find a best synodic period of $6.045$~$\pm$~$0.009$ hours. Note the `dual' 
structure of the periodogram which is expected for an asteroid lightcurve with near
sinusoidal shape.
\textit{Lower panel:} Periodogram for a 5th order Fourier series fit to our $R$-filter time series data.
These higher order fits were performed in order to improve the accuracy of the period.
We find a best-fit synodic rotation period of $6.048$~$\pm$~$0.007$ hours.  
The feature near 8 hours is not physically realistic as it produces a triple-peaked lightcurve. 
The other prominent feature near 12 hours is a harmonic of 6.048.
}
\label{periodogram_o2}
\end{center}
\end{figure}

We applied the method of Harris et al. (\cite{harris}) to determine the 
rotation period.  This method involves fitting an nth-order Fourier 
series to the relative magnitudes, which is then repeated for a wide 
range of periods until the fit residuals are minimized.  The chosen 
range of periods depends on initial inspection of the lightcurves.  
When fitting model lightcurves to the data, we start with low order fits 
to get a feel for where the prominent periodicities reside. 
We then increase the order to refine the dominant periodicity and its associated 
uncertainty, and also the quality of the fit. We stop increasing the order once 
the quality of the fit no longer improves. It was clear from our initial inspection 
of the data that 
the asteroid was rotating in $\sim$6 hours, assuming a double-peaked lightcurve.  
We created a periodogram over the reasonable range of 0--15 hours, which is shown
in the upper panel of Figure \ref{periodogram_o2}.
This periodogram results from 2nd order fits and shows
the location of two dominant periodicities.
As we know the object has a full rotation period of just over 6 hours (assuming a 
double-peaked lightcurve), we adopt the feature at 6.045 hours as the measured synodic period. 
The other prominent minima at 
3.022 hours is simply the result of folding the lightcurve at 
half this frequency resulting in low fit residuals, which is expected for an 
asteroid lightcurve with near sinusoidal shape.
If the lightcurve had a more asymmetric shape, then the periodogram feature 
around 3.022 hours would be much less prominent.

Our final fit, using a 5th order polynomial to refine the period estimate above,
is shown in the lower panel of Figure \ref{periodogram_o2}.  
Prominent minima are seen at approximately 6, 8, and 12 hours. 
The best-fit synodic period is $6.048$~$\pm$~$0.007$ hours.
The feature near 8 hours is not physically realistic as it produces a triple peaked lightcurve. 
The other prominent feature near 12 hours is just a harmonic of 6.048.  

\begin{figure}[t]
\begin{center}
\resizebox{\hsize}{!}{\includegraphics{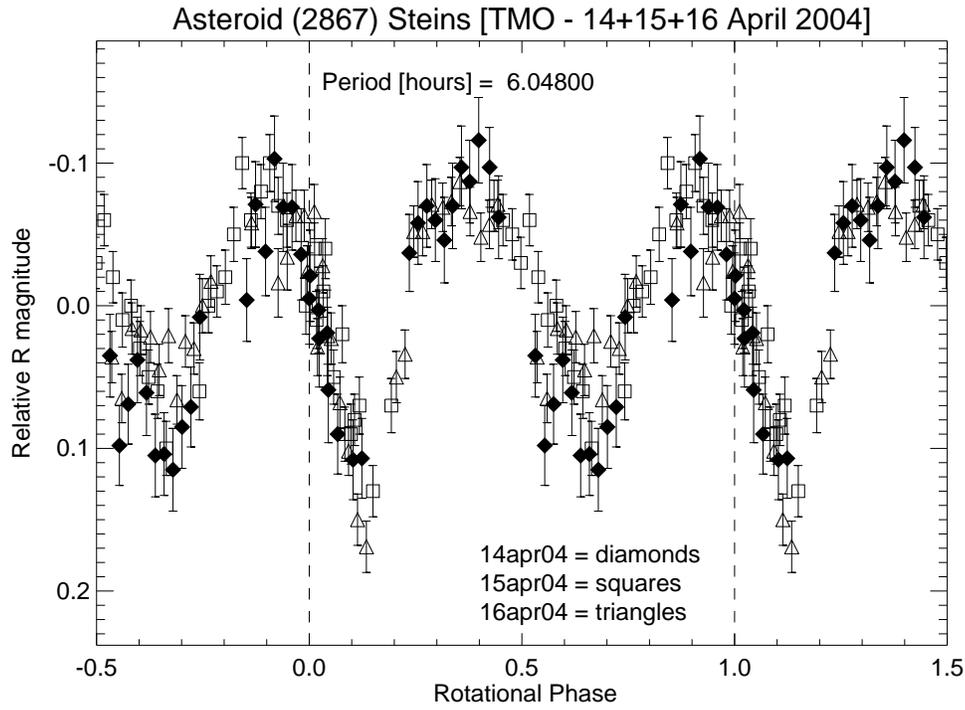}}
\caption[Caption]
{The $R$-filter relative magnitudes phased to the best-fit synodic period of 6.048 hours.
Data from different nights are distinguished by the various symbol types. The repeatability 
from night to night is excellent. The full brightness variation was 
$0.29$~$\pm$~$0.04$ magnitudes.
}
\label{phasecurve}
\end{center}
\end{figure}

In Figure \ref{phasecurve}, we plot the 
relative magnitudes vs rotational phase. The points are folded to the best-fit synodic period of
6.048 hours, and the relative magnitudes for each night have been scaled according to
their nightly averages and then centered around zero, so that small geometry changes are accounted for
in the folding of the data points.  
The double-peaked asymmetric lightcurve covering one full rotation is clearly visible each night.  
Our derived period agrees well with the previous value obtained 
by Hicks et al. (IAUC 8315), who found a period of $6.06$~$\pm$~$0.05$ hours. 

\begin{figure}[t]
\begin{center}
\resizebox{\hsize}{!}{\includegraphics{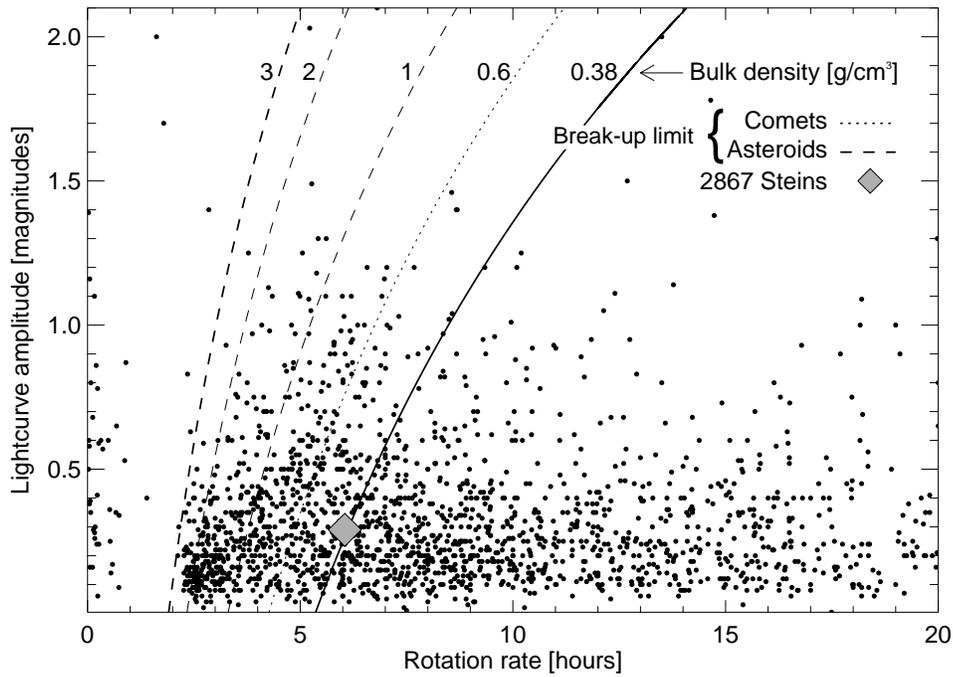}}
\caption[Caption]
{Comparison of our Steins rotational lightcurve parameters
with other asteroidal bodies. The curves are lines of constant
bulk density for a simple centrifugal break-up model (see text).
The rubble-pile breakup limit is shown for the asteroid population
at a density of 3 g/cm$^{3}$. 
The equivalent break-up limit for cometary nuclei at 0.6 g/cm$^{3}$ is also marked.
The location of Steins on this chart indicates that this asteroid
is very typical of asteroidal bodies.
From this model we measure a bulk density lower limit for Steins of 0.38 g/cm$^{3}$.
}
\label{shape_rotation_fig}
\end{center}
\end{figure}

In Figure \ref{shape_rotation_fig} we compare our Steins rotational 
lightcurve parameters with other asteroidal bodies. The overplotted curves are 
lines of constant bulk density for a simple centrifugal break-up model,
and for various density values. 
For a strengthless prolate ellipsoidal body, the critical rotation period $P_{critical}$ - 
beyond which a body will be broken apart by centrifugal forces exceeding self gravity -
can be approximated by
\begin{equation}
\label{density_equation}
P_{critical} \approx \frac{3.3}{\sqrt{\rho}}\sqrt{\frac{a}{b}}
\end{equation}
where $P_{critical}$ is in units of hours, $\rho$ [g/cm$^3$] is the bulk density of the body,
and $a/b$ is the axial ratio derived from the lightcurve amplitude as described in section 
\ref{Size_and_shape} (Pravec \& Harris \cite{pravec_harris}). $P_{critical}$ and
$\rho$ are derived directly from the lightcurve, although the latter is only a lower limit as the
rotation axis orientation is unknown and the derived axial ratio is the projected value.
Also, there is no requirement that the asteroid be rotating at its critical period.
Therefore one can only obtain a lower limit to the bulk density from the lightcurve parameters.

The rubble-pile breakup limit is shown for the asteroid population
at a density of 3 g/cm$^{3}$. The only asteroids that are able to spin 
faster than the $\sim$2 hour spin rate are the so-called monolithic
fast rotators. The equivalent break-up limit for cometary nuclei is also
marked at 0.6 g/cm$^3$ (Lowry and Weissman \cite{lowry}).
One can see that Steins is very typical in terms of rotation period 
and elongation.
We applied this simple break-up model to Steins to place a bulk density lower
limit of 0.38 g/cm$^3$. 
Strictly speaking, the full magnitude range is not yet known, which could affect
this density lower limit determination.


\section{Discussion and Summary}
\label{Summary}

We have analyzed results from photometric observations of the Rosetta
mission flyby target asteroid 2867 Steins, obtained on April 14-16, 2004 (UT)
at the Table Mountain Observatory, California. These data, when combined with 
future data at different observing geometries, 
will be critical for developing a detailed 3-D shape and orientation model (crucial for Rosetta's
science and operations planning), as well as allowing 
surface properties to be investigated in detail prior to the flyby in September 2008.
Our main conclusions are as follows:

\begin{enumerate}

\item We measured a mean apparent $R$-filter magnitude of $16.80$~$\pm$~$0.02$. 
If we assume a phase-slope parameter G of 0.15, as used by Hicks et al. (IAUC 8315),
we obtain a mean absolute $R$-filter magnitude of $12.60$~$\pm$~$0.02$.  This is
slightly brighter than the Hicks et al. measurement but consistent at the 2$\sigma$ level.
If we assume a typical E-type-asteroid G value of 0.40 (the type assigned
to Steins by Barucci et al. \cite{barucci}) we obtain a mean 
absolute $R$-filter magnitude of $12.86$~$\pm$~$0.02$.

\item The observed rotational lightcurve was asymmetric, and the observed full brightness 
range was $0.29$~$\pm$~$0.04$ magnitudes. This implies an axial ratio, $a/b$, of 
$\geq 1.30$~$\pm$~$0.04$.

\item We fitted a 5th order Fourier series to the time-series relative magnitudes and  
found a best-fit synodic rotation period of $6.048$~$\pm$~$0.007$ hours, in good agreement with other observers.
The rotation rate and shape of Steins is very typical for asteroidal bodies.
Using a centrifugal break-up model we determined a bulk density lower
limit of 0.38 g/cm$^3$.

\item 
Assuming a typical S-type albedo of 0.20 (and G=0.15), 
our mean apparent $R$-band magnitude implies a mean effective radius of 
$3.57$~$\pm$~$0.03$~km or semi-axes of $3.42$~$\pm$~$0.03$ and 
$4.06$~$\pm$~$0.03$~km if the full phase-angle-corrected
magnitude range is considered. Assuming a typical E-type albedo of 0.40 (and G=0.40), 
we obtain a smaller mean effective radius of 
$2.24$~$\pm$~$0.02$~km or semi-axes of $2.14$~$\pm$~$0.02$ 
and $2.54$~$\pm$~$0.02$~km. 

\item 
We use the available $R$-filter photometry for a preliminary assessment of the phase-angle
variation of the asteroid's brightness. We combined our mean apparent magnitudes 
with the Hicks et al. data (IAUC 8315) to perform a fit of the photometric phase function 
in terms of the HG formalism. 
The best fit phase function parameters and associated 1$\sigma$ uncertainties are  
G=$0.46^{+0.32}_{-0.20}$, and H=$12.92^{+0.22}_{-0.17}$. Opposition-surge coverage would 
be useful along with observations at large phase angles in order to refine this measurement.

\item Our measured colour indices are: ($V-R$) = $0.58$~$\pm$~$0.03$, 
($R-I$) = $0.44$~$\pm$~$0.03$, and thus ($V-I$) = $1.03$~$\pm$~$0.04$. 
These values agree well with Hicks et al. 
(IAUC 8315) but suggest that if Steins is an E-type asteroid as suggested 
by Barucci et al. (\cite{barucci}), then it is an unusually red E-type object.

\end{enumerate}


\begin{acknowledgements}
We thank the referees for their comments on an earlier 
draft of this paper.
This work was performed in part at the Jet Propulsion Laboratory 
under a contract with NASA and at Queens University, Belfast.  This work 
was funded in part by the NASA Rosetta and Planetary Astronomy Programs.
SCL gratefully acknowledges support from the Leverhulme Trust.
Table Mountain Observatory is operated 
under internal funds from JPL's Science and Technology Management 
Council.  IRAF is distributed by the National Optical Astronomy 
Observatories, which is operated by the Association of Universities 
for Research in Astronomy, Inc. We acknowledge JPL's Horizons online 
ephemeris generator for providing the asteriod's position and rate of 
motion during the observations.

\end{acknowledgements}

\clearpage


\begin{table}
\caption{Orbital elements: 2867 Steins}             	
\label{orbital_elements}      				
\begin{tabular}{l l }        				
\hline\hline                 				
Element & Value$^{\mathrm{a}}$   \\    			
\hline                        				
   semimajor axis & 2.3633 AU  \\      			
   eccentricity & 0.145478  \\
   inclination & 9.9456 degrees  \\
   argument of perihelion & 250.4844 degrees  \\
   longitude of ascending node & 55.5399 degrees  \\
   time of perihelion & 2001 Nov. 04.21279 \\
\hline                                   		
\end{tabular}
\begin{list}{}{}
\item[$^{\mathrm{a}}$] Epoch: 2005-Jan-30.
\end{list}
\end{table}

\begin{table}
\caption{Observing geometry (at 4 hr UT)}             		
\label{observing_geometry}      				
\begin{tabular}{l l l c}        				
\hline\hline                 					
Date & r (AU) &  $\Delta$ (AU) &  Phase angle (deg.)  \\    	
\hline                        					
   14 April 2004 & 2.5601 & 1.7731 & 16.78  \\	   		
   15 April 2004 & 2.5588 & 1.7809 & 17.07  \\
   16 April 2004 & 2.5576 & 1.7890 & 17.35  \\
\hline                                   			
\end{tabular}
\end{table}

\begin{table}
\caption{Apparent $R$-filter magnitudes}             	
\label{table_phot}      				
\begin{tabular}{cccccc}        				
\hline\hline                 				
Date$^{\dagger}$ & $m_R$ $\pm$ $\sigma$ & Date$^{\dagger}$ & $m_R$ $\pm$ $\sigma$ & Date$^{\dagger}$ & $m_R$ $\pm$ $\sigma$\\    
\hline                        				

09.62304  & 16.763 $\pm$ 0.032 &10.62904 &16.811 $\pm$ 0.020 &  11.62060 & 16.808  $\pm$  0.024\\
09.62857  & 16.771 $\pm$ 0.030 &10.63420 &16.771 $\pm$ 0.018 &  11.62598 & 16.790  $\pm$  0.023\\
09.63371  & 16.787 $\pm$ 0.030 &10.63939 &16.801 $\pm$ 0.018 &  11.63113 & 16.761  $\pm$  0.018\\
09.63990  & 16.858 $\pm$ 0.028 &10.65213 &16.901 $\pm$ 0.019 &  11.63612 & 16.761  $\pm$  0.018\\
09.64901  & 16.876 $\pm$ 0.028 &10.65789 &16.891 $\pm$ 0.018 &  11.64205 & 16.758  $\pm$  0.019\\
09.65426  & 16.875 $\pm$ 0.028 &10.66877 &16.941 $\pm$ 0.018 &  11.64699 & 16.796  $\pm$  0.019\\
09.68235  & 16.731 $\pm$ 0.027 &10.67967 &16.881 $\pm$ 0.019 &  11.65202 & 16.847  $\pm$  0.018\\
09.68752  & 16.710 $\pm$ 0.029 &10.74053 &16.741 $\pm$ 0.018 &  11.65723 & 16.892  $\pm$  0.018\\
09.69274  & 16.698 $\pm$ 0.029 &10.74587 &16.751 $\pm$ 0.018 &  11.66232 & 16.926  $\pm$  0.017\\
09.69778  & 16.708 $\pm$ 0.029 &10.75133 &16.761 $\pm$ 0.018 &  11.66766 & 16.974  $\pm$  0.018\\
09.70309  & 16.722 $\pm$ 0.030 &10.75642 &16.781 $\pm$ 0.018 &  11.67288 & 16.993  $\pm$  0.018\\
09.70796  & 16.698 $\pm$ 0.030 &10.76155 &16.751 $\pm$ 0.018 &  11.69066 & 16.874  $\pm$  0.018\\
09.71323  & 16.671 $\pm$ 0.029 &10.76674 &16.791 $\pm$ 0.018 &  11.69583 & 16.858  $\pm$  0.017\\
09.71812  & 16.681 $\pm$ 0.029 &10.77238 &16.821 $\pm$ 0.019 &  11.70094 & 16.772  $\pm$  0.018\\
09.72344  & 16.652 $\pm$ 0.030 &10.77748 &16.811 $\pm$ 0.018 &  11.70624 & 16.772  $\pm$  0.017\\
09.72992  & 16.671 $\pm$ 0.028 &10.78270 &16.841 $\pm$ 0.019 &  11.71160 & 16.753  $\pm$  0.017\\
09.73527  & 16.706 $\pm$ 0.029 &10.78789 &16.861 $\pm$ 0.019 &  11.71813 & 16.755  $\pm$  0.017\\
09.75697  & 16.803 $\pm$ 0.029 &10.79314 &16.871 $\pm$ 0.019 &  11.72325 & 16.751  $\pm$  0.017\\
09.76257  & 16.866 $\pm$ 0.028 &10.79837 &16.911 $\pm$ 0.019 &  11.72834 & 16.737  $\pm$  0.017\\
09.76782  & 16.837 $\pm$ 0.028 &10.81798 &16.871 $\pm$ 0.020 &  11.73433 & 16.758  $\pm$  0.017\\
09.77325  & 16.806 $\pm$ 0.030 &10.82328 &16.811 $\pm$ 0.019 &  11.74071 & 16.776  $\pm$  0.017\\
09.77860  & 16.829 $\pm$ 0.030 &10.82836 &16.801 $\pm$ 0.018 &  11.74583 & 16.767  $\pm$  0.017\\
09.78385  & 16.873 $\pm$ 0.029 &10.83333 &16.791 $\pm$ 0.019 &  11.75075 & 16.753  $\pm$  0.017\\
09.78904  & 16.872 $\pm$ 0.029 &10.83837 &16.761 $\pm$ 0.019 &  11.77404 & 16.860  $\pm$  0.018\\
09.79431  & 16.883 $\pm$ 0.029 &10.84334 &16.711 $\pm$ 0.018 &  11.77997 & 16.889  $\pm$  0.017\\
09.79971  & 16.853 $\pm$ 0.029 &10.84851 &16.751 $\pm$ 0.019 &  11.78600 & 16.840  $\pm$  0.018\\
09.80495  & 16.839 $\pm$ 0.028 &10.85445 &16.731 $\pm$ 0.019 &  11.79130 & 16.841  $\pm$  0.017\\
09.81022  & 16.776 $\pm$ 0.030 &10.85966 &16.711 $\pm$ 0.020 &  11.79699 & 16.846  $\pm$  0.018\\
09.83800  & 16.764 $\pm$ 0.029 &10.86466 &16.741 $\pm$ 0.020 &  11.80215 & 16.869  $\pm$  0.018\\
09.84326  & 16.697 $\pm$ 0.030 &10.86982 &16.751 $\pm$ 0.019 &  11.80769 & 16.845  $\pm$  0.019\\
09.84917  & 16.730 $\pm$ 0.031 &10.88739 &16.821 $\pm$ 0.020 &  11.81269 & 16.890  $\pm$  0.017\\
09.85447  & 16.665 $\pm$ 0.030 &10.89260 &16.771 $\pm$ 0.021 &  11.81770 & 16.849  $\pm$  0.018\\
09.85960  & 16.699 $\pm$ 0.031 &10.89749 &16.861 $\pm$ 0.019 &  11.82274 & 16.854  $\pm$  0.018\\
09.86487  & 16.699 $\pm$ 0.030 &10.90264 &16.831 $\pm$ 0.020 &  11.82770 & 16.824  $\pm$  0.019\\
09.87026  & 16.732 $\pm$ 0.032 &10.90776 &16.901 $\pm$ 0.021 &  11.83280 & 16.807  $\pm$  0.018\\
09.87556  & 16.747 $\pm$ 0.032 &10.91278 &16.881 $\pm$ 0.020 &  11.85688 & 16.766  $\pm$  0.018\\
09.88091  & 16.791 $\pm$ 0.034 &  --	 &        --         &  11.88991 & 16.800  $\pm$  0.018\\
09.88637  & 16.827 $\pm$ 0.037 &  --	 &        --         &  11.89593 & 16.853  $\pm$  0.020\\
  --	  &        --          &  --	 &        --         &  11.90191 & 16.845  $\pm$  0.020\\
\hline                                   
\end{tabular}\\
$\dagger$ - Light-time corrected mid-exposure-JD minus 2453100
\end{table}

\clearpage


\begin{table}
\caption{$BVRI$ Colours for Asteroid 2867 Steins}         
\label{table:4}      				
\begin{tabular}{c c c c}        			
\hline\hline                 			
Colour &  This work &$\overline{R}$$^{\dagger}$ &  Hicks et al.~(IAUC 8315) \\    
\hline                        			
   $V-R$ & $0.58 \pm 0.03$ &$16.91 \pm 0.02$	& $0.51 \pm 0.03$  \\
   $R-I$ & $0.44 \pm 0.03$ &$16.92 \pm 0.02$	& $0.44 \pm 0.03$  \\
   $B-R$ & $-$  		&		& $1.311 \pm 0.03$ \\
\hline                                   	
\end{tabular}\\
$\dagger$ - Corresponding average apparent $R$ magnitude (see text)
\end{table}


\end{document}